\documentclass{PoS}

\usepackage[draft,inline,nomargin]{fixme}

\usepackage{physics}
\usepackage{booktabs}

\usepackage{floatrow}

\usepackage{cite}

\title{$I=3/2$ $N\pi$ scattering and the $\Delta(1232)$ resonance on 
$N_{\mathrm{f}}=2+1$ CLS ensembles using the stochastic LapH method}

\ShortTitle{$N\pi$ scattering and the $\Delta(1232)$ resonance}

\author{\speaker{Christian Walther Andersen}$^a$ John Bulava$^a$, Ben 
H\"orz$^{b}$ and Colin Morningstar$^{c}$\\
	\llap{$^a$}CP$^3$-Origins \& Dept. of Mathematics and Computer Science, 
	University 
	of Southern Denmark\\
	Campusvej 55, 5230 Odense M, Denmark\\
	\llap{$^b$}Nuclear Science Division, Lawrence Berkeley National Laboratory\\
	Berkeley, CA 94720, USA\\
	\llap{$^c$}Dept. of Physics, Carnegie Mellon University\\
	Pittsburgh, Pennsylvania 15213, USA\\
	
	E-mail: \email{cwandersen@imada.sdu.dk}}

\abstract{
Calculations of the elastic $I=\frac{3}{2}$  nucleon-pion scattering 
phase shifts on two lattice QCD ensembles with $m_\pi=200\mathrm{MeV}$ and 
$280\mathrm{MeV}$ are presented. The ensembles both employ $N_\mathrm{f} = 2+1$ 
Wilson 
clover fermions. We determine the $\Delta(1232)$ resonance parameters from a 
finite volume scattering analysis. In one study the single partial 
wave simplification is employed to compute the $p$-wave amplitude
while in the other we treat 
the partial wave mixing between $s$- and $p$-wave contributions.
Fitting our data to a Breit-Wigner resonance model we find $m_\Delta/m_\pi = 
7.13(9)$ and $4.75(5)$ on the two ensembles respectively, showing that for a 
lighter quark mass the resonance mass moves from near the $N\pi$ threshold to 
near the $N\pi\pi$ threshold, in agreement with experiment.
}

\FullConference{37th International Symposium on Lattice Field Theory - Lattice2019\\
		16-22 June 2019\\
		Wuhan, China}

\begin{document}
	
\newcommand{\ecm}{E_\mathrm{cm}}
\newcommand{\qcm}{\vb{q}_\mathrm{cm}^2}

\section{Introduction}

In this work we present a determination of elastic $N \pi$ scattering 
amplitudes. Since the calculations are performed using
simulations 
of QCD on a 
Euclidean lattice a direct determination of the scattering observables is not 
possible \cite{Maiani:1990ca}. A common way to circumvent this problem makes 
use of the fact that the discrete, interacting energy levels in a finite 
spatial volume are 
shifted from their non-interacting values by an amount that can be related to 
the scattering matrix. This relation 
was first described for scattering between two identical, spinless particles 
with total zero momentum by L\"uscher \cite{Luscher:1990ux}. This result has 
since been extended and generalized with advances relevant for this work found 
in Refs.
\cite{Gottlieb:1995dk,Fu:2011xz,Luu:2011ep,Leskovec:2012gb,Gockeler:2012yj,
	Hansen:2012tf,Briceno:2014oea,Morningstar:2017spu},
allowing scattering studies of increasingly 
impressive precision, in particular in the meson-meson sector.
When introducing baryons into the calculations one has to deal 
with a more 
severe 
signal-to-noise problem, increased computational cost and a more involved 
analysis dealing with particles with differing spin, so 
comprehensive studies 
of resonant 
meson-baryon and baryon-baryon scattering are still lacking.

\section{Methods}

The two gauge field ensembles used in this work were generated by the CLS 
consortium \cite{Bruno:2016plf,Bali:2016umi}. Both simulations employ 
$N_{\mathrm{f}}=2+1$ dynamical Wilson clover fermions and both have open 
boundary conditions in the time direction. All interpolating operators are 
kept at a minimum distance of $t_{\mathrm{bnd}}$ from the boundary, where 
$t_{\mathrm{bnd}}m_\pi = 2$. The ensemble details are listed in Tab. 
\ref{tab:ensembles}. Results on the N401 ensemble have been 
published in 
Ref. \cite{Andersen:2017una}.

\begin{table}[]
	\centering
	\begin{tabular}{@{}llllllll@{}}
		\toprule
		ID   & $\beta$ & $a(\mathrm{fm)}$ & $\left(\tfrac{L}{a}\right)^3 \times 
		\tfrac{T}{a}$   & 
		$m_\pi, m_K 
		(\mathrm{MeV})$ & $N_{\mathrm{conf}}$ & $N_{t_0}$ & $N_{\mathrm{ev}}$\\ 
		\midrule
		N401 & $3.46$  & $0.0763$        & $48^3 \times 128$ & $280, 
		460$                 & 275                 & 2    &  320 \\
		D200 & $3.55$  & $0.0643$        & $64^3 \times 128$ & $200, 
		480$                 & 559                 & 2    &  448   \\ 
		\bottomrule
	\end{tabular}
	\caption{The ensembles used in this calculation 
		\cite{Bruno:2014jqa}. The table specifies the gauge 
		coupling $\beta$, 
		the lattice constant $a$ \cite{Bruno:2016plf}, the extent of the 
		lattices in lattice units, 
		masses of the pseudoscalar particles, the number of configurations used 
		in 
		the analysis $N_\mathrm{conf}$, the number of source times $N_{t_0}$ 
		and the number of eigenvectors used in the smearing $N_\mathrm{ev}$.}
	\label{tab:ensembles}
\end{table}

The required finite volume energy levels are extracted from 
correlation 
functions of operators with $\Delta(1232)$ quantum numbers including 
$N\pi$-operators. 
In order to compute correlation functions of 
multihadron operators with definite momentum we employ all-to-all quark 
propagators, 
which are efficiently handled with the stochastic LapH method 
\cite{Morningstar:2011ka}. In this framework, the quark 
propagator is projected into a 
lower dimensional subspace constructed from $N_{\mathrm{ev}}$ eigenvectors of 
the stout 
smeared \cite{Morningstar:2003gk} gauge-covariant 3-D lattice Laplace operator. 
In this way the color and space indices of the quark propagator are converted 
to eigenvector indices. Converting back to color and space indices results in 
a spatially smeared quark field that retains all symmetries of the original 
unsmeared field.

The quark propagator is stochastically estimated in the LapH subspace spanned 
by time, spin and eigenvectors. The number of eigenvectors $N_{\mathrm{ev}}$ 
used in the smearing  can be seen in Tab. \ref{tab:ensembles}.
We use dilution \cite{Wilcox:1999ab,Foley:2005ac} 
in time, spin 
and eigenvector indices 
to reduce the variance of the stochastic estimation. The dilution scheme is 
explained in Ref. \cite{Andersen:2017una} for the N401 lattice.
Using that notation the dilution scheme for the D200 data is 
$(\mathrm{TF},\mathrm{SF},\mathrm{LI}8)_F$, 
$(\mathrm{TI}8,\mathrm{SF},\mathrm{LI}8)_R$, re-using the solutions 
to the 
Dirac equation from Ref. \cite{Andersen:2018mau}.
To further increase statistics we average over two source times, 
all equivalent total momenta $\vb{P}$ and all irrep rows $\lambda$.

Two of the practical problems encountered in resonant meson-baryon 
scattering lattice calculations are the 
proliferation of Wick contractions that need to be evaluated and the extra 
annihilation type diagrams in correlation functions between single baryon and 
meson-baryon interpolating operators. Although not employed in the present 
work, one way of dealing with the increasing number of Wick contractions is 
explained in Ref. \cite{Horz:2019rrn}.

In order to determine as many energy levels as possible the fact that
our interpolating operators have different overlaps on the lowest lying 
states in the spectrum is exploited\cite{Michael:1985ne,Luscher:1990ck}. In 
practice, we 
calculate the energy spectrum for various total momenta $\vb{P}$ and in various 
irreducible representations $\Lambda$ of the little group of $\vb{P}$. For 
each pair of $(\Lambda,\vb{P})$ of interest a matrix of correlation functions 
$C_{ij}(t) = \expval{\hat{O}_i(t) \hat{O}^\dagger_j(0)}$ is computed for 
a number of 
operators projected to the given irrep and total momentum. The operator basis 
consists of 1-2 single-site $\Delta$ operators and 2-7 $N\pi$ operators 
depending on the irrep and total momentum. We then solve the generalized 
eigenvalue problem (GEVP)
\begin{equation}\label{eq:gevp}
C(t_d)v_n(t_d,t_0) = \lambda_n C(t_0) v_n(t_d,t_0),
\end{equation}
and the correlators between ``optimal'' interpolators for the 
$n$'th
state can then be found by rotating the correlator matrix by the eigenvectors:
\begin{equation}\label{eq:opt_corr}
\hat{C}_n(t) = (v_n(t_0,t_d), C(t)v_n(t_0,t_d)).
\end{equation}
We make sure that the extracted energies are stable under 
variation of $t_0, 
t_d$ and the operator basis.

This optimal correlator is expected to decay exponentially with the energy of 
the $n$'th state with overlap on the operator basis. We now form the quantity
\begin{equation}\label{eq:ratio}
R_n(t) = \frac{\hat{C}_n(t)}{C_\pi(\vb{p}_{\pi,n}^2,t)C_N(\vb{p}_{N,n}^2,t)},
\end{equation} 
where $C_\pi(\vb{p}_{\pi,n}^2,t)$ and $C_N(\vb{p}_{N,n}^2,t)$ are the 
correlators of the interpolating pion and nucleon respectively. 
For each finite volume level we chose $\vb{p}_{\pi,n}$ and $\vb{p}_{N,n}$ 
corresponding to a nearby non-interacting level.
$R(t)$ is then expected to decay exponentially with the energy shift from the 
non-interacting energies, and thus we fit $R(t)$ to the ansatz
\begin{equation}\label{eq:ratio_ansatz}
R_n(t)=Ae^{-\Delta E_n t},
\end{equation}
and reconstruct the total energy $E_n$ from $\Delta E_n$ and the measured 
values of the pion and nucleon energies.

Given the finite volume energies, elastic 2-to-2 scattering amplitudes can be 
computed using the determinant equation
\begin{equation}\label{eq:det_cond}
\det(\hat{K}^{-1} - B^{(\vb{P})}) = 0.
\end{equation}
For a given total momentum $\vb{P}$ and irreducible representation $\Lambda$ of 
the little group of $\vb{P}$, $\hat{K}$ and $B^{(\vb{P})}$ are matrices in 
total angular momentum $J$, total orbital angular momentum $L$, total spin $S$ 
and occurrence number $n$. $\hat{K}$ is related to the usual 
$K$-matrix by
\begin{equation}\label{eq:k_khat}
\hat{K}^{-1}_{L'S';LS} = 
q_{\mathrm{cm}}^{L+L'+1}K_{L'S';LS}^{-1},
\end{equation}
and is diagonal in $J$. The box matrix $B$ encodes the reduced symmetries of 
the finite volume 
for a given irrep. It is a known matrix of functions of $\ecm$ and is diagonal 
in $S$ and $n$ but dense in the other quantum numbers. All 
$B$-matrix elements required here are given in Ref. \cite{Morningstar:2017spu}.

In this study we are interested in $N\pi$-scattering with isospin 
$I=\frac{3}{2}$. The most prominent feature of this system is the $p$-wave 
$\Delta(1232)$ resonance ($J^\eta = \frac{3}{2}^+$ where $\eta$ is the parity), 
which is the focus of the rest of this work. The 
first published lattice calculation of the resonant phase shifts of this system 
appeared in Ref. \cite{Andersen:2017una}, but other preliminary 
work can be 
found in Refs. \cite{Paul:2018yev,Mohler:2012nh,VerduciValentina2014Psil}.
The $\Delta$-baryons decay almost exclusively to $N\pi$ states \cite{PGD}, 
so we need only worry about a single open channel.

In order to numerically evaluate Eq. 
(\ref{eq:det_cond}) the 
matrices must be 
truncated at some $L_\mathrm{max}$. Since the $\Delta$ resonance occurs in 
$N\pi$ 
scattering with $L=1$ setting $L_\mathrm{max} \ge 1$ is required. We 
check the impact of the $d$-wave explicitly by varying $L_\mathrm{max}$ from 1 
to 2.

In the irreps $(\Lambda,\vb{P}^2) = \left\lbrace (H_g,0), (G_2,1), (F_1,3), 
(F_2,3), (G_2,4) \right\rbrace$ the $B$-matrix elements corresponding to 
$J=\frac{1}{2}$ and/or $L=0$ are identical to 0. 
This means that if $L_\mathrm{max}=1$ there 
is a 1-to-1 correspondence between a measured energy level and a $p$-wave 
scattering phase shift.
We also measure the energy spectrum in the irreps $(\Lambda,\vb{P}^2) = 
\left\lbrace (G_{1g},0), (G_{1u},0), (H_{u},0), (G_1,1), (G,2), (G,3), (G_1,4) 
\right\rbrace$. In these irreps we have to take partial wave mixing 
into account.
However for $N\pi$ scattering the $\hat{K}$-matrix is 
fully diagonal in $J$ and $L$, so we can write
\begin{equation}\label{eq:diag_k_matrix}
\hat{K}^{-1} = \mathrm{diag}\left( (\hat{K}^{-1})_{\frac{1}{2},0}, \;
(\hat{K}^{-1})_{\frac{1}{2},1}, \; (\hat{K}^{-1})_{\frac{3}{2},1} \right),
\end{equation}
with subscripts denoting $(J,L)$, and so only three elements of the $\hat{K}$ 
matrix need to be parameterized.

An important limitation of this formalism is that Eq. (\ref{eq:det_cond}) only 
holds below any relevant three-particle thresholds. For this system the first 
such three-particle state is $N\pi\pi$, meaning that any energy above $m_N 
+ 2m_\pi$ is excluded from the scattering analysis.

\section{Results}

The energies are determined by fitting the optimized correlators 
to the ansatz in Eq. 
(\ref{eq:ratio_ansatz}) from some minimum time separation $t_{\mathrm{min}}$ to 
a 
fixed maximum time separation of $t_{\mathrm{max}}=25a$. 
The value of $t_{\mathrm{min}}$ is chosen large enough that the statistical 
error on the 
fitted energy is larger than the systematic error from the excited 
state 
contamination. We determine 6 and 26 energy levels in the elastic region on 
the 
N401 and D200 ensembles respectively.

%


The measured energies can now be inserted into Eq. (\ref{eq:det_cond}) to 
constrain the $\hat{K}$-matrix elements. We do this by 
parameterizing the elements of 
the $\hat{K}$-matrix and then fitting those parameters using a correlated 
$\chi^2$ fit. The residuals of the fitting procedure are taken to be 
\cite{Morningstar:2017spu}
\begin{equation}\label{eq:residuals}
\Omega(\mu,A) = \frac{\det(A)}{\det(\left[\mu^2+AA^\dagger\right]^{1/2})}
\end{equation}
with $A=\hat{K}^{-1}-B^{(\vb{P})}$ and $\mu = 5$.
Since a Breit-Wigner resonance in the $J=\frac{3}{2}^+$ $p$-wave is expected we 
fit
\begin{equation}\label{eq:k_mat_param_p}
\left(\hat{K}^{-1}\right)_{\frac{3}{2},1} = 
\left(\frac{m_\Delta^2}{m_\pi^2} - 
\frac{\ecm^2}{m_\pi^2}\right)\frac{6\pi\ecm}{g^2m_\pi},
\end{equation}
with the fit parameters $m_\Delta/m_\pi$ being the resonance mass 
in units 
of the pion mass and $g$ which is related to the resonance width. For assessing 
$s$- and $d$-wave contributions we use a truncated effective range expansion 
with just one fit parameter per $\hat{K}$-matrix element:
\begin{equation}\label{eq:k_mat_param_sd}
\left(\hat{K}^{-1}\right)_{J,L} = \frac{1}{m_\pi^{2L+1}a_{J,L}}, 
\qquad (J,L) \neq \left(\tfrac{3}{2},1\right).
\end{equation}

The results of the fits are shown in Tab. \ref{tab:results}. For the 
N401 ensemble we 
only measure energies in the irreps where the $p$-wave is the lowest 
contributing partial wave and so make no attempt to include an $s$-wave 
parameterization of the $\hat{K}$-matrix. We additionally perform a 
fit including a
$d$-wave parameterization and an extra energy level in the 
$(\Lambda,\vb{P}^2)=(H_u,0)$, where the $d$-wave is the lowest contributing 
partial wave. Doing this had no significant impact on the $p$-wave scattering 
parameters.
On the D200 we also study the impact of including 
irreps where the $s$-wave is present. Tab. \ref{tab:results} shows that this 
reduces the statistical error on the resonance mass by a factor of 2 compared 
to the fit using only $p$-wave levels and improves 
the quality of the fit significantly.
We also see that including a $d$-wave parameterization 
actually shifts the resonance mass slightly outside of its statistical error. 
However, no extra $d$-wave energy levels were 
included here since they were all above the inelastic threshold, and it 
should also be noted that the $\chi^2/\mathrm{d.o.f.}$ is rather low, 
suggesting some over fitting.

\floatbox[{\capbeside\thisfloatsetup{capbesideposition={left,center},
		capbesidewidth=0.35\textwidth}}]{figure}[\FBwidth]
{\caption{\emph{Bottom:} The energies of all 26 states included in the 
		scattering analysis. Colored errorbars show states in irreps where the 
		$p$-wave is the lowest contributing partial wave with one color per 
		irrep (see legend). 
		\emph{Middle and top:} In irreps where we apply the single-partial wave 
		approximation the value of the 
		$\hat{K}$-matrix element and $\delta_{\frac{3}{2},1}$ respectively is 
		shown with 
		errorbars from bootstrapping. The fit (including all 26 energies) is 
		shown with dotted lines.}
	\label{fig:comb}}
{\includegraphics[width=0.5\textwidth]{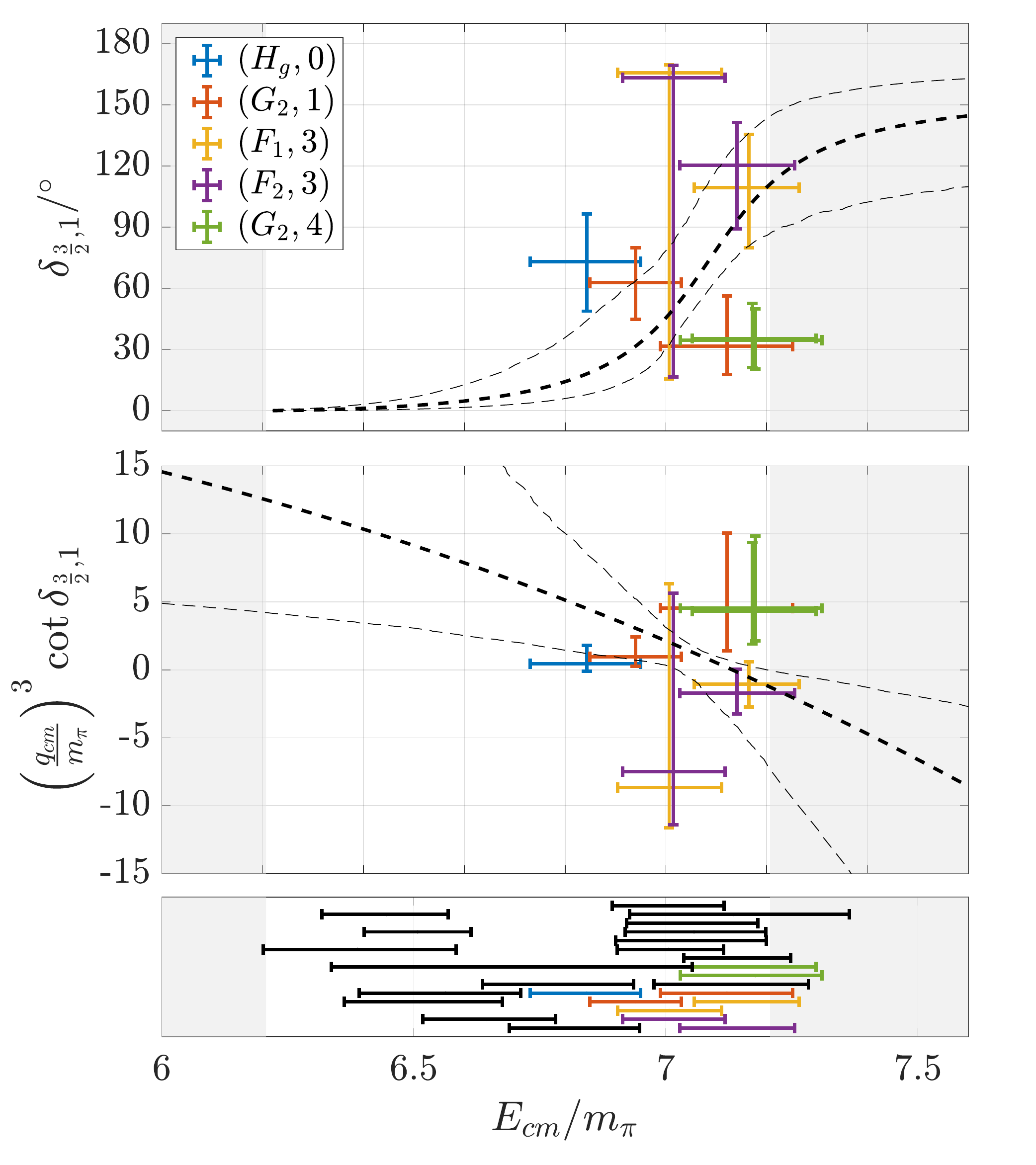}}

Fig. \ref{fig:comb} shows the result of the calculation on the D200 
lattice. All energies determined and included in the analysis are shown in the 
bottom panel while the upper panels show the value of 
$\left(\hat{K}^{-1}\right)_{\tfrac{3}{2},1} = 
\left(\tfrac{q_\mathrm{cm}}{m_\pi}\right)^3 \cot \delta_{\tfrac{3}{2},1}$ and 
$\delta_{\frac{3}{2},1}$ for the irreps with a trivial determinant condition. 
The dotted lines show the fit to $s$- and $p$-wave including all 26 energy 
levels.

\begin{table}[]
	\centering
	\begin{tabular}{@{}llllll@{}}
		\toprule
		ID   & $L$'s & $N_E$ & $\frac{m_\Delta}{m_\pi}$ & $g$     
		& 
		$\chi^2/\mathrm{d.o.f.}$ \\ \midrule
		N401 & 1 & 6     & $4.75(5)$                & $19(5)$ & 
		$1.1$                   \\
		N401 & $1,2$ & 7     & $4.73(6)$            & 
		$19(7)$ & 
		$4.2$                   \\
		D200 & $1$ & 9    & $7.2(2)$                & $18(11)$ & 
		$1.7$ 					\\
		& $0,1$ & 26    & $7.13(9)$            & $11(6)$ & 
		$0.8$                    \\ 
		& $0,1,2$ & 26    & $6.88(19)$         & $25(17)$ & 
		$0.4$                    \\\bottomrule
	\end{tabular}
	\caption{Result of the scattering analyses. $L$'s indicates the partial 
		waves parameterized in the $\hat{K}$-matrix, $N_E$ is the number of 
		energy 
		levels included and the fitting parameters are explained in the text.}
	\label{tab:results}
\end{table}

\section{Conclusion}

We have presented a preliminary calculation of 
$N\pi$ scattering phase shifts complementing already published results 
\cite{Andersen:2017una}. The two ensembles have different $m_\pi$, lattice 
spacing and physical 
volume. Assuming a Breit-Wigner fit-form, we get a good estimate of 
the $J=\frac{3}{2}$ $p$-wave 
resonance mass although the width is still not determined precisely.
Including energy levels in irreps which mix partial waves helps 
to 
constrain the $p$-wave scattering parameters despite the mixing.
While we expect the systematic errors from the finite volume and 
lattice spacing to be small we postpone a quantitative assessment of these 
effects.
In the future we plan to further increase statistics on both 
ensembles using 
improved estimators 
with more dilution projectors and more gauge configurations to obtain more 
precise constraints on the scattering amplitude parameters.

%
%

\bibliographystyle{JHEP}
\bibliography{cwa.bib}


\end{document}